\documentclass[aps,prl,preprint,showpacs,superscriptaddress]{revtex4-1}  
\usepackage{latexsym,amsmath,amssymb,amstext}
\usepackage{slashed}
\usepackage{epsfig}
\usepackage{float,xcolor}
\usepackage{graphicx}
\usepackage{dcolumn}

\newcommand{\tr}{{\rm tr\ }}

\newcommand{\be}{\begin{equation}}
\newcommand{\ee}{\end{equation}}
\newcommand{\bea}{\begin{eqnarray}}
\newcommand{\eea}{\end{eqnarray}}
\newcommand\bef{\begin{figure}}
\newcommand\eef[1]{\label{fg:#1}\end{figure}}
\newcommand\beq{\begin{equation}}
\newcommand\eeq[1]{\label{#1}\end{equation}}
\newcommand\beqa{\begin{eqnarray}}
\newcommand\eeqa[1]{\label{#1}\end{eqnarray}}
\newcommand\bet{\begin{table}}
\newcommand\eet[1]{\label{tb:#1}\end{table}}

\newcommand\fgn[1]{Figure \ref{fg:#1}}
\newcommand\eqn[1]{Eq.\ (\ref{#1})}

\newcommand\tbn[1]{Table \ref{tb:#1}}

\usepackage[paperwidth=210mm,paperheight=297mm,centering,hmargin=2cm,vmargin=2.5cm]{geometry}

\begin{document}

\widetext

\title{Monopole scaling dimension using Monte Carlo}
\author{Nikhil\ \surname{Karthik}}
\email{nkarthik@bnl.gov}
\affiliation{Physics Department, Brookhaven National Laboratory, Upton, New York 11973-5000, USA}

\begin{abstract}
We present a viable Monte Carlo determination of the scaling
dimensions $\Delta_Q$ of flux $Q$ Abelian monopoles through finite-size
scaling analysis of the free energy to introduce the background
field of classical Dirac monopole-antimonopole pair at critical
points of three-dimensional lattice theories.  We validate the
method in free fermion theory, and by verifying the  particle-vortex
duality between the monopole scaling dimension at the inverse-XY
fixed point and the charge scaling dimension at the XY fixed point.
At the $O(2)$ Wilson-Fisher fixed point,
we determine the critical exponents $\Delta_1= 0.13(2)$, $\Delta_2=0.29(1)$
and $\Delta_3=0.47(2)$, which we find to be proportional to the
finite-size critical spectrum of monopoles on square
torus.
\end{abstract}

\maketitle

\section{Introduction}
Classification of all possible relevant
operator deformation of critical points is central to understanding
critical phenomena, and the field theories which either have their
continuum limits at the critical point or flow to the fixed point
in the infra-red limit~\cite{RevModPhys.51.659}.  This involves a
computation of scaling dimension of operators at the fixed point.
In three dimensions, in addition to the usual local operators which
are analytic functions of the field or spin variables in the
underlying theory, there are also the disorder operators whose
action is to introduce non-trivial winding of the the field or spin
variables about the insertion point~\cite{Murthy:1989ps,Borokhov:2002ib}.
In the case of theories with $U(1)$ symmetry, these are the magnetic
monopole operators which introduce $2\pi Q$ total magnetic flux on
spheres enclosing them, and their scaling dimensions $\Delta_Q$ are
a new set of critical exponents.

The relevance of monopole terms in the renormalization group flow
could drive quantum field theories to different long distance
behaviors, a well known example being the Abelian non-compact and
compact QED$_3$~\cite{Polyakov:1975rs}, coupled to small number of
massless fermion flavors~\cite{Pufu:2013eda,Pufu:2013vpa}.  In
addition to serving as a possible scale-inducing deformation that
can be added to a fixed point, the monopoles are one of the actors
in the three dimensional particle-vortex dualities in various forms
(e.g.,\
~\cite{Peskin:1977kp,Seiberg:2016gmd,Karch:2016sxi,Metlitski:2015eka,Mross:2015idy}).
These dualities map particles charged under $U(1)$ of one theory
to monopoles of another theory, with the two theories in many cases
being tuned to their critical points.  The most basic three-dimensional
duality is the mathematical correspondence between the Villain form
of the XY model at zero electric charge and the gauged XY model at
zero temperature but non-zero electric charge~\cite{Peskin:1977kp}.
However, many recent particle-vortex dualities are well motivated
but nevertheless conjectural (e.g.,~\cite{Seiberg:2016gmd,Karch:2016sxi}).
Thus, it is imperative that one should be able to compute the
critical exponents of monopole operators using standard
Monte Carlo methods, to go hand in hand with such recent theoretical
developments and also to complement the advancements in bootstrap
methodology in finding scaling
dimensions~\cite{Poland:2016chs}.

The monopole operators are non-local in terms of the fundamental fields,
but behave as local operators~\cite{Borokhov:2002ib}. This
follows from the state-operator correspondence, in which by construction, 
the monopole operators are the local primary operators at the origin
to which the ground states of a CFT in $S^2\times \mathbb{R}$ with net fluxes $2\pi
Q$ over $S^2$, are mapped onto~\cite{Metlitski:2008dw,Dyer:2013fja}. 
Denoting such a primary monopole operator as
$M_Q(x)$, its scaling dimension $\Delta_Q$ at a critical point is
determined from its power-law behavior:
\beq
\left\langle M_Q(x) M_{-Q}(y)\right\rangle \propto \frac{1}{|x-y|^{2\Delta_Q}}.
\eeq{scaling}
In spite of the simplicity of the definition, the actual construction
of a monopole operator itself is subtle in $\mathbb{R}^3$, making
them notoriously difficult to study using the standard Monte Carlo
methods~\cite{Dyer:2013fja}.  In most cases, the ab initio understanding
of monopole operators proceed on a case by case basis, wherein one
maps the monopole operator to trivial local operators in a different
theory related by an established duality or in the same universality
class~\cite{Block:2013gpa,Sreejith:2015ria,PhysRevLett.111.087203}.

The motivation for the current work is to use a Monte Carlo method
for finding $\Delta_Q$, that generalizes to various systems without
appealing to properties special to any lattice model, and demonstrate
that it works.  We do so by coupling  the static background $U(1)$
field from a monopole and antimonopole separated by a non-zero
distance, to the conserved currents of critical lattice theories
and we measure the free energy required to do so. This technique
was first introduced in~\cite{Murthy:1989ps} for the case of ${\rm
CP}^\infty$ model where the partition function can be exactly
computed, and now routinely used in analytic perturbative computations
of $\Delta_Q$ using the state-operator
correspondence~\cite{Pufu:2013eda,Dyer:2015zha,Dyer:2013fja,Metlitski:2008dw,Murthy:1989ps,Pufu:2013vpa}.
From the asymptotic scaling of this free energy with the distance
between the monopole and antimonopole, which is kept proportional
to the lattice size itself, we determine the monopole scaling
dimension. The underlying assumption is that the introduction of a
flux $Q$ Dirac monopole background field leads to dominant contributions
from the same configurations that would contribute to the constrained
path integral resulting from the insertion of the scaling operator
$M_Q$, not just in the limit of infinite number of species, $N$,
of spins or matter fields but for any number $N$.

\section{Method}
We consider simple fermion and spin systems with $U(1)$ global
symmetry in $L^3$ periodic lattice.  The $U(1)$ symmetry can be
gauged by coupling the spins to dynamical gauge fields $a_\mu(\mathbf{x})$
which are defined on the links connecting the lattice site $\mathbf{x}$
to $\mathbf{x}+\mathbf{\hat\mu}$.  In addition to the dynamical
gauge fields, one can couple the conserved currents of the systems
to external background fields ${\mathbf {\cal A}}$ in order to
construct ${\cal A}$ dependent partition function and effective
action, $Z({\cal A})$ and $F({\cal A})=-\log Z({\cal A})$ respectively.
In the present work, we set ${\mathbf {\cal A}}$ to be a superposition
of gauge fields from a monopole at $\mathbf{r}_0$ and an antimonopole
at $\mathbf{r}'_0$, which are separated by a distance
$r=|\mathbf{r}_0-\mathbf{r}'_0|$, and compute the response of the
system to change in $r$. That is, the superposed field $\mathbf{{\cal
A}}^{\rm
Q\overline{Q}}(\mathbf{r};r)=\mathbf{A}^{Q}(\mathbf{r};\mathbf{r}_0)-\mathbf{A}^{Q}(\mathbf{r};\mathbf{r}^\prime_0)$,
where $\mathbf{A}^{Q}(\mathbf{r};\mathbf{r}_0)$ is the classical,
scale-covariant field at a point
$\mathbf{r}=(x,y,z)$ from a Dirac monopole of magnetic charge $Q\in \mathbb{Z}$
at $\mathbf{r}_0=(x_0,y_0,z_0)$ (c.f.,~\cite{Balian:2005joa}) :
\beq
\mathbf{A}^Q(\mathbf{r};\mathbf{r}_0) = \frac{Q}{2} \frac{(\mathbf{r}-\mathbf{r_0})\times \hat{\mathbf{z}}}{|\mathbf{r}-\mathbf{r_0}|(|\mathbf{r}-\mathbf{r_0}|-(z-z_0))}.
\eeq{dmono}
Instead of naively discretizing the continuum solution on the
lattice, we compute the gauge transporters ${\tilde {\cal
A}}^{Q\overline{Q}}_\mu(\mathbf{r};r)=\int_\mathbf{r}^\mathbf{r+\hat\mu}
dx_\mu {\cal A}^{Q\overline{Q}}_\mu(\mathbf{x};r)$ exactly, and
couple it to the conserved current of the models. On the lattice,
we separate $Q$ and $\overline{Q}$ by $\mathbf{r_0}-\mathbf{r_0'}=r
\hat{\mathbf{z}}$, such that the center of mass of the $Q\overline{Q}$-pair
is at the center of the lattice. For this choice, $r$ is the length
of the Dirac string that runs between $Q$ and $\overline{Q}$. Since
the lattice is periodic, we take $\mathbf{A}^{Q}(\mathbf{x})$ to
be \eqref{dmono} if $x_\mu \le L$ and force periodicity otherwise.
Any jump in $\mathbf{A}^{Q}(\mathbf{x})$ itself at $x_\mu = L$ is
proportional to $1/L$, and any contribution of such a jump to the
effective action will be suppressed further by a surface to volume
factor in the thermodynamic limit.

The monopole-antimonopole correlator in the background field method is  simply 
\beq
G(r) \equiv \frac{Z\left(\tilde{{\cal A}}^{Q\overline{Q}}(r)\right)}{Z(0)}.
\eeq{moncorr}
One can compute such a difference in free energies with and without
the background field through Monte Carlo simulation by introducing
auxiliary variables in the action~\cite{Kajantie:1998zn}.  Let
$\zeta\in [0,Q]$ be such an auxiliary variable, then
\begin{equation}
F\left(\tilde{{\cal A}}^{Q\overline{Q}}(r)\right)-F(0)=\int_0^Q d\zeta W(\zeta);\  W(\zeta;r)\equiv\frac{\partial}{\partial \zeta}F(\zeta \tilde{{\cal A}}^{1\overline{1}}) .
\end{equation}
The quantity $W$ is a measurement that can be made in Monte Carlo
simulation of $Z(\zeta A^{1\overline{1}})$ theory. Henceforth, we
refer to the above difference in free energies simply as $F_Q(r)$.

We assume that the monopole-antimonopole correlator
at the critical point of lattice theories is a scaling function,
$G(r,L,\xi_L) \sim r^{-2\Delta_Q} g\left(\frac{r}{\xi_L},\frac{r}{L}\right)$,
where $\xi_L$ is the finite correlation length in the finite system
which we will grow linearly with $L$ at the critical point. There
could be corrections from finite size scaling from subleading
$L^{-\omega}$ terms.  A finite-size scaling method to determine the
exponents $\Delta_Q$ is to consider the correlation functions at
$r=\rho L$ for a fixed fraction $\rho$ in different lattice sizes~\cite{Banerjee:2017fcx}.
In such a case, $G(\rho L,L,\xi_L)\sim L^{-2\Delta_Q}\left(1+
O\left(L^{-\omega}\right)\right)$. Therefore, we look for the following
finite size behavior of free energy with a leading logarithmic term
\beq
F_Q(\rho L)=f_0(\rho,Q)+2\Delta_Q\log(L)+\frac{f_1(\rho,Q)}{L^\omega},
\eeq{fssdimmon}
in order to extract $\Delta_Q$.  The value of $\omega$ for local
operators is known in 3d XY and Ising models to be close to
0.8~\cite{Hasenbusch:1999cc,Campostrini:2006ms,Hasenbusch:2000ph}.  Due
to our lack of knowledge about such scaling-violation exponent for
background insertions, we simply use an analytic $\omega=1$ which
empirically accounts for any corrections to scaling in the volumes
we study. Throughout this paper, we set $\rho=r/L=1/4$ for finite-size
scaling studies.

\section{Systems} 
To serve as a sanity check, we study the partition
function $Z_F({\cal A})$ for a single two-component free Dirac
fermion coupled to the external field ${\cal A}^{Q\overline{Q}}$.
We use the two-component Wilson-Dirac fermion for this purpose, in
which case, $Z_F=\det\left(\slashed{D}_W({\cal A})\right)$ with
$\slashed{D}_W$ being the Wilson-Dirac operator.  Then, we study the
chargeless limit and the zero temperature limit of the lattice
superconductor model with the action~\cite{Dasgupta:1981zz} 
\beqa
S&=&-\beta \sum_x \sum_{\mu=1}^3 \cos\left(\nabla_\mu\theta(x)+e
a_\mu(x)+\tilde{{\cal A}}^{Q\overline{Q}}_\mu(x)\right)\cr &
&+\frac{1}{2}\sum_x\sum_{\mu>\nu=1}^3 \left(\nabla_\mu a_\nu(x)-\nabla_\nu
a_\mu(x)\right)^2, 
\eeqa{sxy} 
where $\nabla_\mu f(x)=f(x+\hat\mu)-f(x)$.  The first $e=0$ case
is the XY model, whose critical point at $\beta_c= 0.4541652$ lies
in the $O(2)$ universality
class~\cite{Campostrini:2000iw,Hasenbusch:1999cc}.  The second
$\beta\to\infty$ limit corresponds to the frozen superconductor
(FZS) model whose critical point~\cite{Neuhaus:2002fp,Dasgupta:1981zz}
at $e^2_c=13.148997$ is in the inverse-XY universality
class~\cite{Dasgupta:1981zz}. In the FZS limit, the arguement of
cosine is forced to take the values $2\pi n_\mu$ for integer valued
$n_\mu$~\cite{Neuhaus:2002fp}.  The FZS action becomes
\beq 
S = \frac{2\pi^2}{e^2} \sum_x\sum_{\mu>\nu=1}^3 \left(\nabla_\mu
n_\nu(x)-\nabla_\nu n_\mu(x)-\frac{\tilde{\cal{F}}^{Q\overline{Q}}_{\mu\nu}(x)}{2\pi}\right)^2,
\eeq{fzsaction} 
with $\tilde{{\cal F}}^{Q\overline{Q}}_{\mu\nu}(x)=\nabla_\mu \tilde{{\cal
A}}^{Q\overline{Q}}_\nu(x)-\nabla_\nu \tilde{{\cal A}}^{Q\overline{Q}}_\mu(x)$.  An
exact particle-vortex duality mapping between the XY model and FZS
model was worked out by Peskin~\cite{Peskin:1977kp}.  In this
duality, the charge-$Q$ operators $e^{iQ\theta(x)}$ maps onto the
monopole operators $M_Q(x)$ in the FZS model.  The critical exponents
for the charge-$Q$ operators at XY fixed point are well
known~\cite{Banerjee:2017fcx,PhysRevB.84.125136}.

\bef
\centering
\includegraphics[scale=1.0]{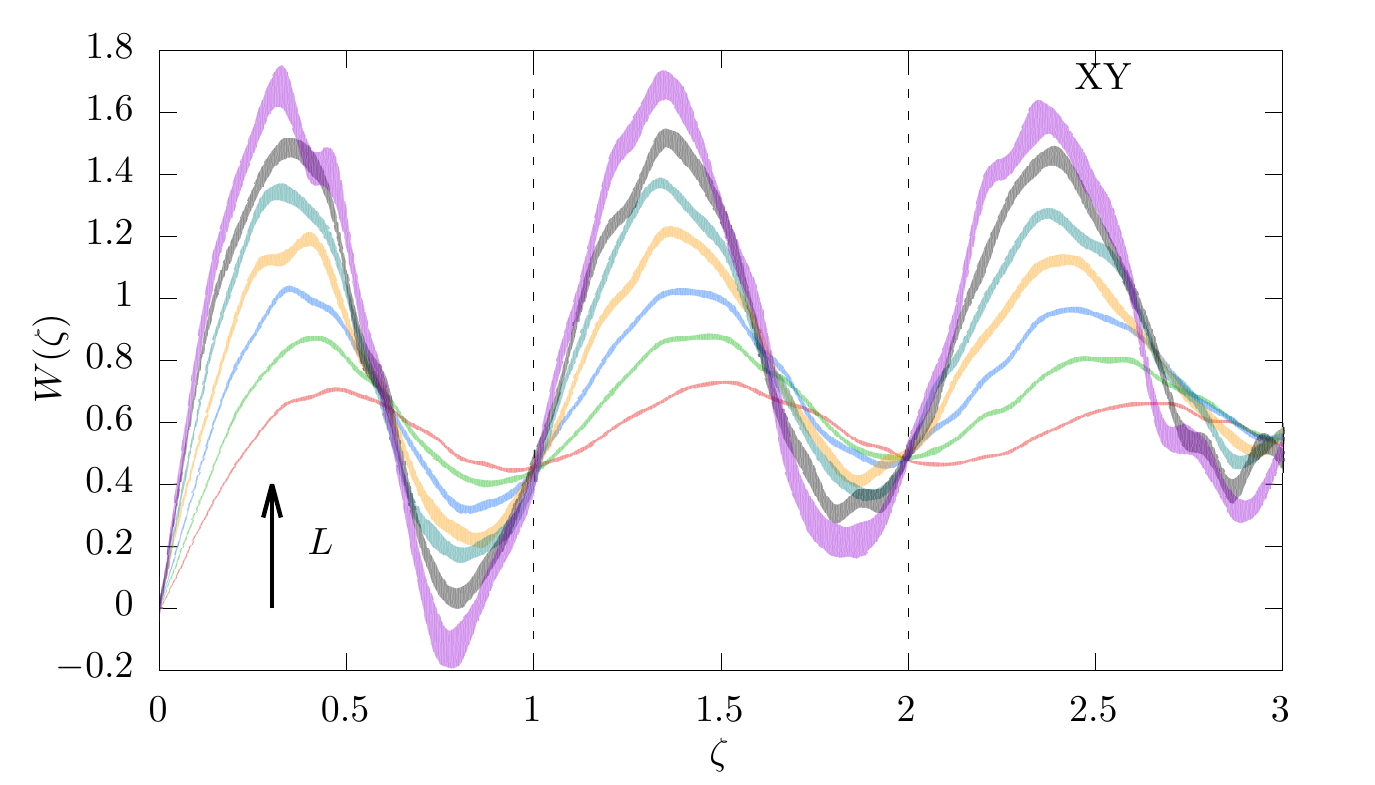}
\caption{
The function $W(\zeta)$ is shown in the range 0 to 3 for
monopole-antimonopole separation $r=L/4$ in the critical XY model.
The different colored curves are the interpolation curves of
$W(\zeta)$ from different $L$. Along the direction of the arrow,
$L=12,16,20,24,28,32$ and 36.
}
\eef{wplotferm}

\begin{figure*}
\vskip -0.15cm
\centering
\includegraphics[scale=0.8]{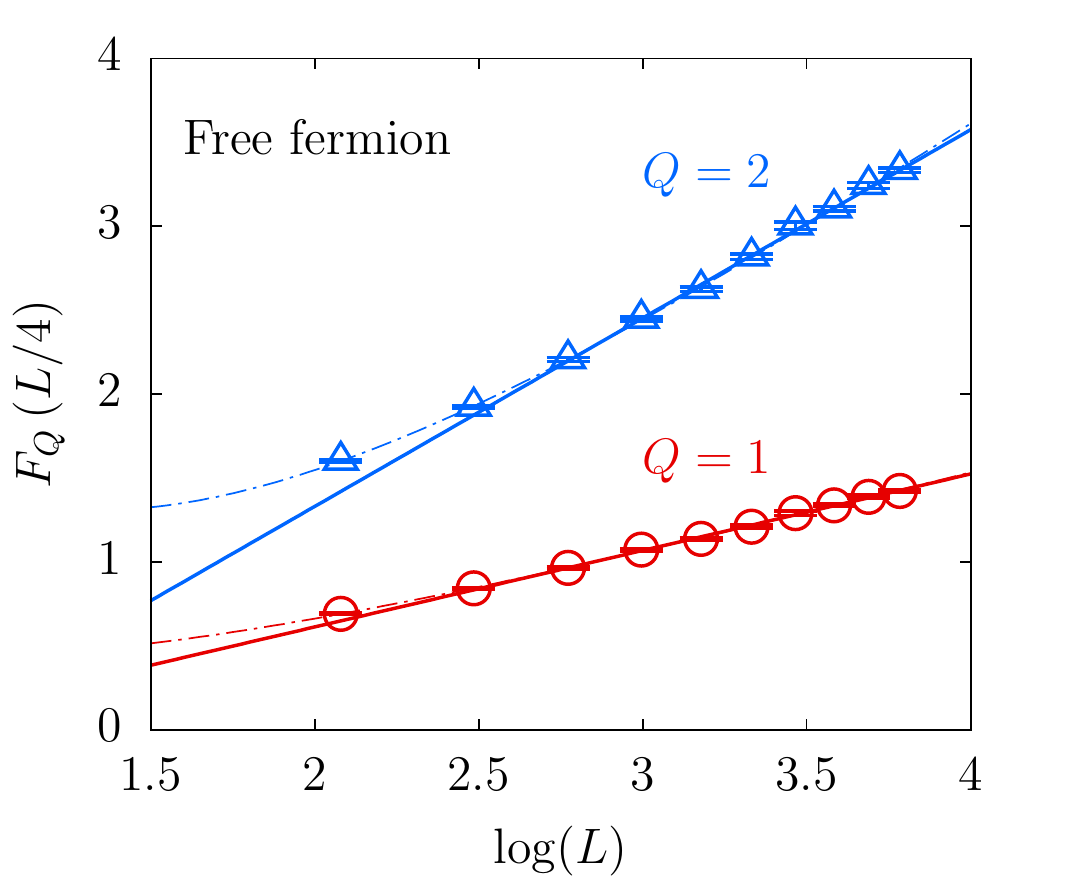}
\includegraphics[scale=0.8]{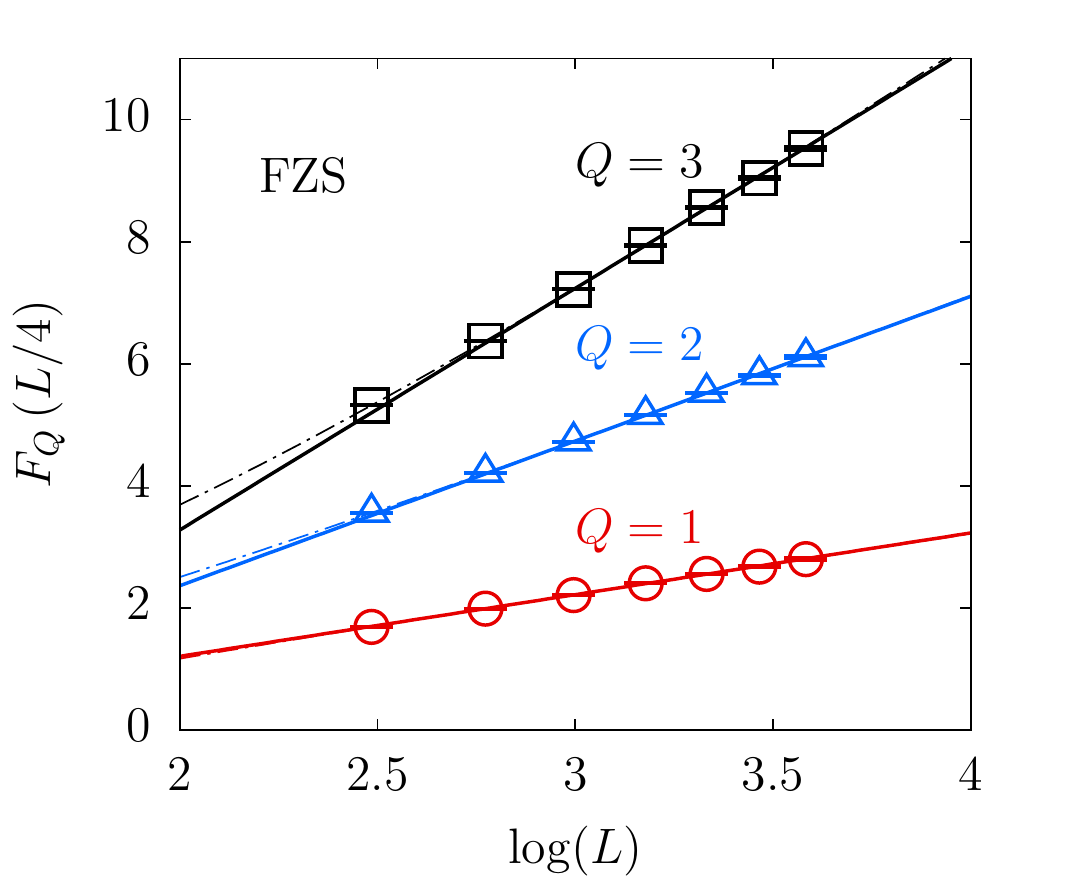}
\includegraphics[scale=0.8]{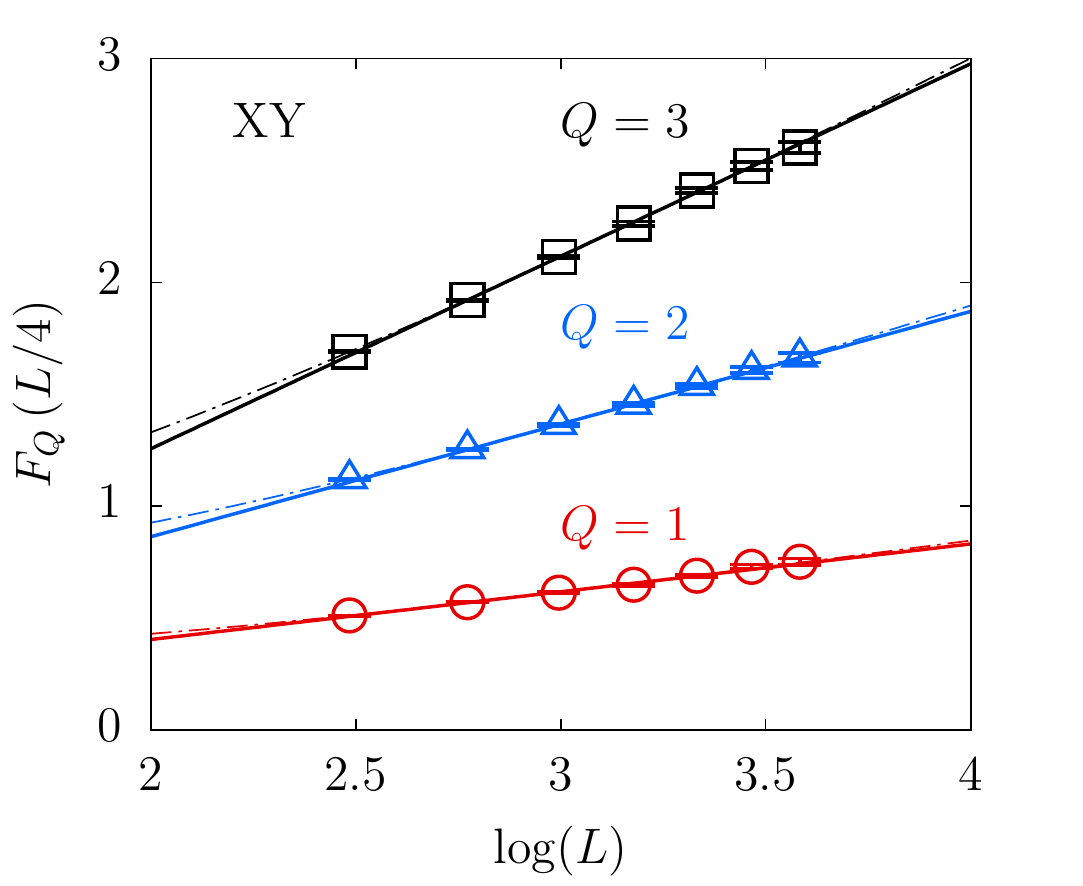}
\vskip -0.15cm
\caption{
The free energy of monopole-antimonopole background field insertion
is shown as a function of $\log(L)$ for fixed $\rho=r/L=1/4$. The
top panel is for free Wilson-Dirac fermion, the middle panel
for the critical frozen superconductor model and the bottom panel
for the critical XY model.  The curves are fits to the data.
}
\label{fg:monodim}
\end{figure*}

\bet
\centering
\begin{tabular}{|c|c||l|l||l|}
\hline
Model & $Q$ & \multicolumn{2}{c|}{$\Delta_Q$} & Expectation \\
\hline
Free fermion & 1 & 0.227(4) &  0.253(8) & 0.265\quad (e)\\
             & 2 & 0.561(8) &  0.66(1)  & 0.673\quad (e)\\
\hline
    & 1 & 0.51(1) & 0.48(2) & 0.516(3)\ (d)\\
FZS & 2 & 1.18(1) & 1.23(2) & 1.238(5)\ (d)\\
    & 3 & 1.97(1) & 2.15(4) & 2.116(6)\ (d)\\
\hline
   & 1 & 0.107(4) & 0.13(2) & 0.065\quad (a)\\
XY & 2 & 0.252(3) & 0.29(1) & 0.159\quad (a)\\
   & 3 & 0.429(5) & 0.47(2) & 0.272\quad (a)\\
\hline
\end{tabular}
\vskip -0.15cm
\caption{Table of estimated scaling dimensions for free Wilson
fermion, critical FZS model and critical XY model.  The third and
the fourth columns tabulate the fit values of $\Delta_Q$ with and
without a $1/L$ scaling correction term respectively.  The fifth column is the expected
values; the entries marked (e) are exact results, those marked (d)
are inferred from particle-vortex duality, while those marked (a)
are expectations based on large-$N$ calculations.  }
\eet{dimtab}

\section{Monopole critical exponents}
For the finite-size scaling study, we used periodic $L^3$ lattices
for $L=12,16,20,24,28,32$ and 36.  For each fixed values of $L$ and
$Q$, the different $\zeta$ corresponds to independent Monte Carlo
simulations. We used 48 different values of $\zeta$ from $0$ to $3$
for each $L^3$ lattice in order to study $Q=1,2,3$.  We simulated
the XY model at the critical point $\beta_c$ using Hybrid Monte
Carlo (HMC) global updates~\cite{Duane:1987de}.  We made about
$5.10^6$ measurements in all our lattice sizes. For the FZS model,
we used single-hit Metropolis algorithm and made $10^8$ such updates.
Error estimates were made using  block Jack-knife to account for
autocorrelations.  For the free Wilson-Dirac fermion, we evaluated
$W(\zeta)=-\tr\left(\slashed{D}^{-1}_W\slashed{D}^\prime_W\right)$
stochastically using $10^4$ random vectors.  We also tuned the
Wilson mass on $A^{1\overline{1}}$ background so that the lattice
fermion is massless~\cite{Karthik:2015sgq}.

In \fgn{wplotferm}, we show $W(\zeta;r)$ as a function of $\zeta$
for the critical XY model. In order to obtain the curves, we
interpolated the equally spaced Monte-Carlo data points for
$W(\zeta;r)$ using cubic spline. The different curves correspond
to different $L$ at fixed $\rho=1/4$. In order to obtain the free
energy for $Q$-monopole, we integrate the splines from 0 up to $Q$.
We obtain similar such curves for the free fermion as well as the
critical FZS model. For all such cases, we observe distinct
oscillations in $W(\zeta;r)$ of period ${\cal O}(1)$, and the curves
corresponding to fixed $\rho$ approximately intersect each other
close to integer values of $\zeta$.  Due to the charge conjugation
symmetry, $W(\zeta;r)$ is odd in $\zeta$, and in our numerical
simulations we do find $W(0;r)=0$ is satisfied well within error
bars, serving as a check.  At present, we lack a theoretical
understanding of such curves which could help extrapolate the results
to larger $Q$.

In \fgn{monodim}, we show the behavior of free energy $F_Q(r)$ at
fixed $\rho=1/4$ as a function of $\log(L)$.  The three panels from
top to bottom correspond to free Wilson-Fermion, critical FZS and critical
XY models respectively.  The symbols in the plots are the actual
Monte Carlo data.  The most important observation in this paper is
the clear presence of $\log(L)$ behavior in the background field
method and also the onset on this $\log(L)$ behavior for computationally
accessible values of $L$.  The curves are our $\log(L)$ fits to the
data; the solid curve is the straight line fit including just a
$\log(L)$ term using data from $L>12$ lattices, while the dashed
curves include any $1/L$ corrections to the free energy in addition
to the dominant $\log(L)$ term.  In all the cases, the $\chi^2/{\rm
DOF}< 2$ for the fits. 

In the case of free continuum fermion, the values of $\Delta_Q$ are
known exactly by computations of the Casimir energy of free fermions
on $S^2$ with constant flux over it~\cite{Borokhov:2002ib,Pufu:2013eda}.
These values for free fermions are tabulated in \tbn{dimtab} along
with the values of $\Delta_Q$ extracted from fits to the data. There
is about 15\% systematic dependence on the kind of fit. With a $1/L$
correction term included in the fit, the free energy from all $L$
are well described by the fit (\fgn{monodim}),  and the corresponding
fit values of $\Delta_Q$ agree quite well with the analytical results
from free continuum Dirac fermion. While serving as a check on the
method, it is also a fascinating check on the universality of the
nontrivial monopole critical exponent itself as it is determined
using a lattice fermion which only lies in the same universality
class as the free continuum fermion.

The middle panel of \fgn{monodim} shows the result for the critical
FZS model. From the corresponding entries in \tbn{dimtab}, an
excellent agreement with the charge scaling dimensions in $O(2)$
fixed point is seen. This is expected from the exact particle-vortex
duality~\cite{Peskin:1977kp}.  However, it is an important check
that the background field method leads to the same critical exponents
as those of the primary monopole operators that enter the duality, 
thereby supporting our assumption.

Having demonstrated the method in two different cases where the
values of $\Delta_Q$ are available from other means, we apply the
method to $O(2)$ fixed point. There are no dynamical gauge fields
in this case, however one can still insert an external monopole
operator that creates $U(1)$ vortices in $\theta$
fields~\cite{Pufu:2013eda}.  The $\log(L)$ dependence of free energy
data and the fits are shown in the bottom panel of \fgn{monodim},
and we have tabulated the fit values of $\Delta_Q$ in \tbn{dimtab}.
There is about 10\% systematic dependence of
$\Delta_Q$ on the type of fit.  Next to our estimates of $\Delta_Q$,
we also tabulate the expected values from $1/N$ extrapolation of
$\Delta_Q$ calculated in $O(2N)$ fixed points for
large-$N$~\cite{Pufu:2013eda}. The estimated values are about twice
the extrapolated values, perhaps indicating non-negligible
nonpertubative corrections to $\Delta_Q$ for smaller $N$. However,
the conclusion that monopole operators with $Q=1,2,3$ are relevant
($\Delta_Q < 3$) at $O(2)$ fixed point still remains true.

\section{Finite-size spectrum}
\begin{figure}
\vskip -0.15cm
\centering
\includegraphics[scale=0.75]{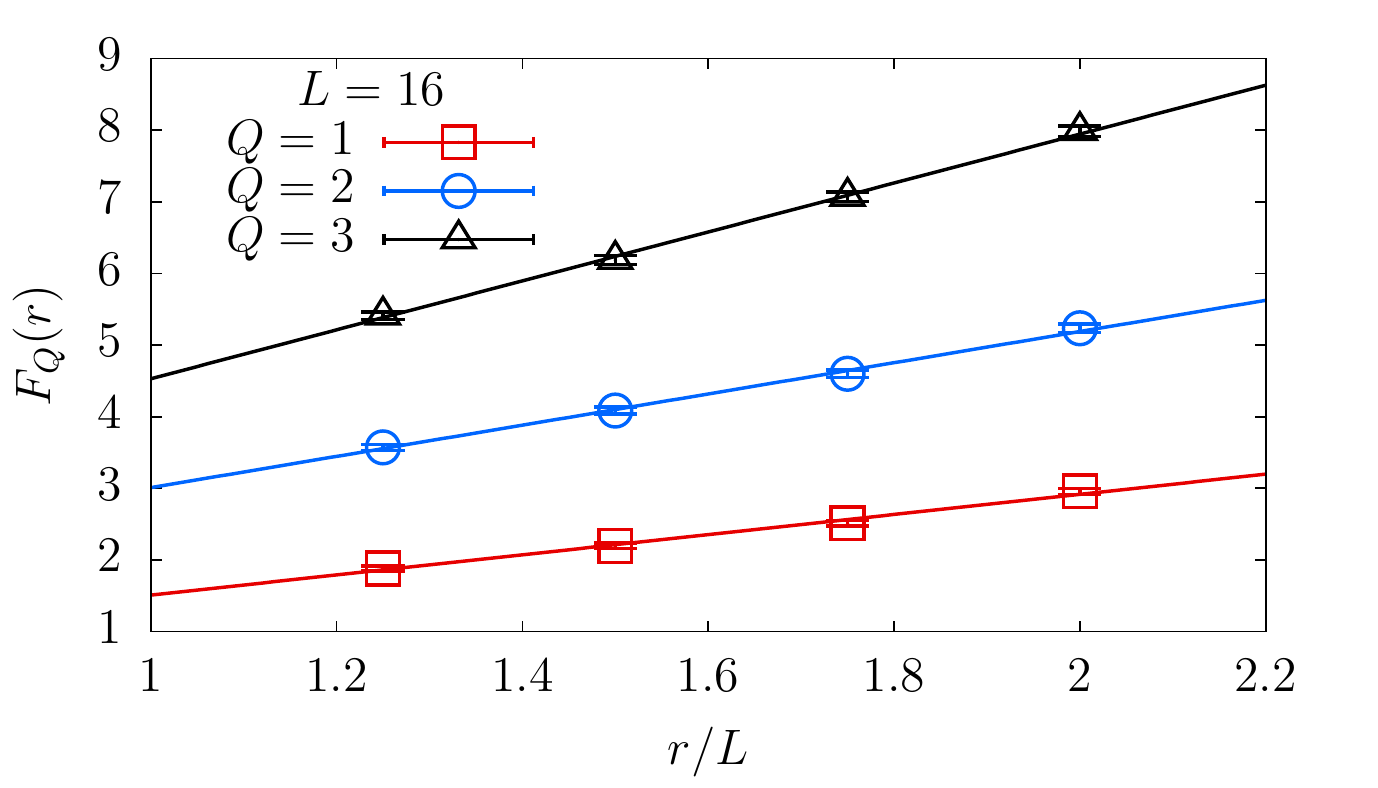}

\includegraphics[scale=0.75]{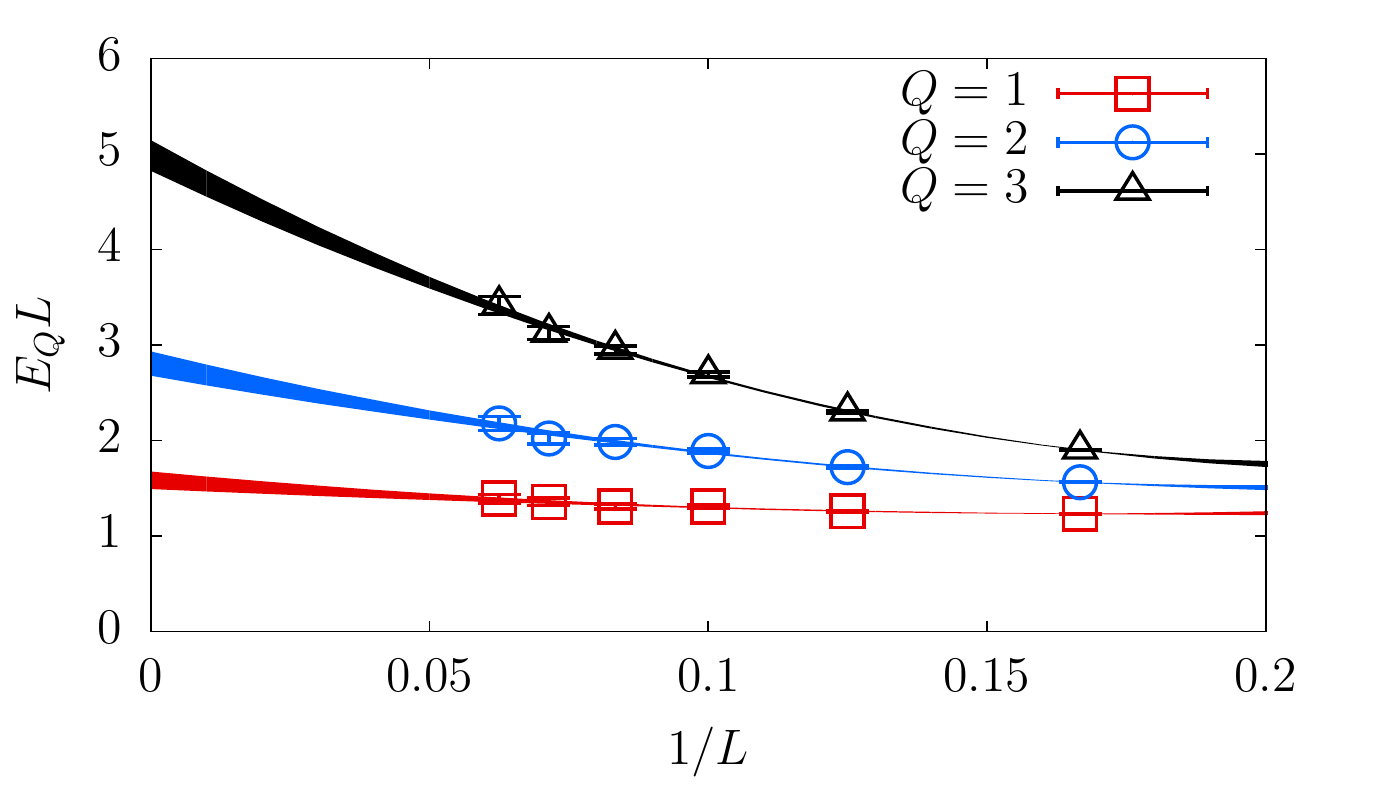}
\vskip -0.3cm
\caption{
Spectrum $E_Q$ of monopoles on torus as determined on $L^2\times
4L$ lattice is shown for the critical XY model. The top panel shows
the linear dependence of free energy with the monopole-antimonopole separation $r$, for $r > L$. The bottom panel
shows the extrapolation of $E_Q L$ to the thermodynamic limit.  }
\label{fg:spectrum}
\end{figure}
Recently, the universal features in the finite-size critical spectrum
of operators have been of
interest~\cite{Thomson:2016ttt,Schuler:2016kuw,Whitsitt:2017ocl}.
Here, we provide a computation of critical spectrum of Dirac monopole
at the $O(2)$ fixed point of XY model.  For this, we use $L^2\times 4L$ lattices
as an approximation for $T^2\times \mathbb{R}$, and compute the
torus finite-size spectrum $E_Q L$ from the slope of a linear
increase in $F_Q(r)$ with $r/L$, for $ L < r < 2 L$ in the thermodynamic
limit $L\to\infty$. The free energy corresponding to such an
`exponential decay' of the monopole-antimonopole correlator is shown
for XY model in the top panel of \fgn{spectrum} on $16^2\times 64$
lattice for $r/L > 1$. In the bottom panel, we show the thermodynamic
limit of the extracted $E_Q L$ using quadratic polynomials in $1/L$.
The existence of the thermodynamic limit of $E_Q L$ is noteworthy
and indicates that the extracted spectrum indeed is that of a
critical theory.  We find $\sqrt{4\pi}\Delta_Q/(E_Q L)$ to be
$0.29(5),0.37(2),0.33(2)$ for $Q=1,2,3$ respectively, indicating a
near proportionality between $\Delta_Q$ and $E_Q$ starting from
small values of $Q$.  Such a ratio for the charge-$Q$ operators,
$\exp(iQ\theta)$, in the XY model was shown to be about 1 even for
small $Q$~\cite{Banerjee:2017fcx,Hellerman:2015nra}.

\section{Discussion}
We demonstrated the effectiveness of a rather straight-forward
numerical implementation of the background field method in determining
the monopole scaling dimensions, as applied to both fermionic and
bosonic critical lattice theories.  We chose simple theories here
in order to test the feasibility of the approach. The successful
application of the method in demonstrating an exact particle-vortex
duality~\cite{Peskin:1977kp} provides ample motivation to apply the
method to recently conjectured particle-vortex dualities.  It would
also be interesting to repeat this computation for the monopole
scaling dimension in the infra-red fixed point of non-compact QED$_3$
with $N$ flavors of massless Dirac fermions to determine the critical
$N$ where monopoles become irrelevant, and check if it matches with
the critical $N$ for compact QED$_3$.  The near proportionality of
$\Delta_Q$ with the torus spectrum $E_Q$ also suggests there could
be universality in the monopole finite-size spectrum similar to the
findings in~\cite{Thomson:2016ttt,Schuler:2016kuw,Whitsitt:2017ocl}.

\begin{acknowledgments}
I would like to thank R. Narayanan for extensive discussions and
critical reading of the manuscript.  I thank D. Banerjee,
S.  Chandrasekharan, R. Dandekar, S. Minwalla, R. Pisarski, and S.
Pufu for useful discussions.  The computations presented in the
paper were carried out using the BC cluster at Fermilab under a
USQCD type-C project.  I thank the Nuclear Theory
Group at BNL for supporting my research.  I acknowledge support by
the U.S. Department of Energy under contract No. DE-SC0012704.
\end{acknowledgments}

\bibliography{biblio}

\clearpage
\newpage
\pagestyle{empty}


\appendix

\section{Monopole background on the lattice}
We use periodic $L_x\times L_y\times L_z$ lattice; for finite-size
scaling studies, $L_x=L_y=L_z=L$, while for extracting torus spectrum,
$L_x=L_y=L$ and $L_z=4 L$, and we use even values of $L$.  Let us
arbitrarily choose a point on the periodic lattice as the origin
which has coordinates as $(1,1,1)$. With respect to this origin,
consider a monopole of magnetic charge $Q$ at $\mathbf{r_0}=(x_0,y_0,z_0)$.
The Dirac monopole background is
\begin{equation}
\mathbf{A}^Q(\mathbf{r};\mathbf{r}_0) = \frac{Q}{2} \frac{(\mathbf{r}-\mathbf{r_0})\times \hat{\mathbf{z}}}{|\mathbf{r}-\mathbf{r_0}|(|\mathbf{r}-\mathbf{r_0}|-(z-z_0))}.
\end{equation}
These field variables are the parallel transporters which live on
the links of the lattice. Therefore, the link variables
connecting point $\mathbf{n}=(x,y,z)$ to $\mathbf{n+\hat\mu}$ is
\begin{equation}
\tilde{\mathbf{A}}^Q_\mu(\mathbf{n};\mathbf{r}_0)=\int_\mathbf{n}^\mathbf{n+\hat\mu} dx_\mu A^Q_\mu(x;\mathbf{r}_0).
\end{equation}
Doing the above integrals, we get
\begin{eqnarray}
\tilde{A}^Q_1(x,y,z;\mathbf{r}_0)&=&\frac{Q}{2}\frac{y-y_0}{|y-y_0|}\big{[}f_1(x-x_0+1,y-y_0,z-z_0)\cr && -f_1(x-x_0,y-y_0,z-z_0)\big{]},\cr
\tilde{A}^Q_2(x,y,z;\mathbf{r}_0)&=&-\frac{Q}{2}\frac{x-x_0}{|x-x_0|}\big{[}f_2(x-x_0,y-y_0+1,z_0)\cr && -f_2(x_0,y-y_0,z-z_0)\big{]},\cr
\tilde{A}^Q_3(x,y,z;\mathbf{r})&=&0,
\end{eqnarray}
where
\begin{eqnarray}
f_1(x,y,z)&=&\tan^{-1}\left(\frac{x}{|y|}\right)+\tan^{-1}\left(\frac{x z}{|y|\sqrt{x^2+y^2+z^2}}\right),\cr
f_2(x,y,z)&=&\tan^{-1}\left(\frac{y}{|x|}\right)+\tan^{-1}\left(\frac{y z}{|y|\sqrt{x^2+y^2+z^2}}\right).\cr &&
\end{eqnarray}
On a periodic lattice it is not possible to have a single monopole.
So, we consider the background field to be the superposition of the
fields due to monopole at position $\mathbf{r}_0$ and an anti-monopole
at position $\mathbf{r}^\prime_0$, which we place in the dual
lattice, in such a way that the pair is almost at the `center' of the periodic
lattice with respect to the coordinate system set by the arbitrary
choice of the origin:
\beqa
\mathbf{r}_0=\begin{cases} \left(\frac{L+1}{2},\frac{L+1}{2},\frac{L+r+1}{2}\right)\quad\text{even $r$} \\
                            \left(\frac{L+1}{2},\frac{L+1}{2},\frac{L+r}{2}\right)\quad\text{odd $r$},
             \end{cases} \cr
\mathbf{r}^\prime_0=\begin{cases} \left(\frac{L+1}{2},\frac{L+1}{2},\frac{L-r+1}{2}\right)\quad\text{even $r$} \\
                            \left(\frac{L+1}{2},\frac{L+1}{2},\frac{L-r}{2}\right)\quad\text{odd $r$}.
             \end{cases}
\eeqa{position}
The superposed field from the monopole-antimonopole pair is
\beq
\tilde{{\cal A}}^{Q\overline{Q}}_\mu(x,y,z;r)=\tilde{A}^Q_\mu(x,y,z;\mathbf{r}_0)-\tilde{A}^Q_\mu(x,y,z;\mathbf{r}^\prime_0),
\eeq{supa}
for $1\le x \le L_x$, $1\le y \le L_y$ and $1\le z \le L_z$.  For
$x_\mu\to x_\mu+ L_\mu$, we  force periodic boundary conditions on
$\tilde{{\cal A}}^{Q\overline{Q}}$.
\bef
\centering
\includegraphics[scale=0.35]{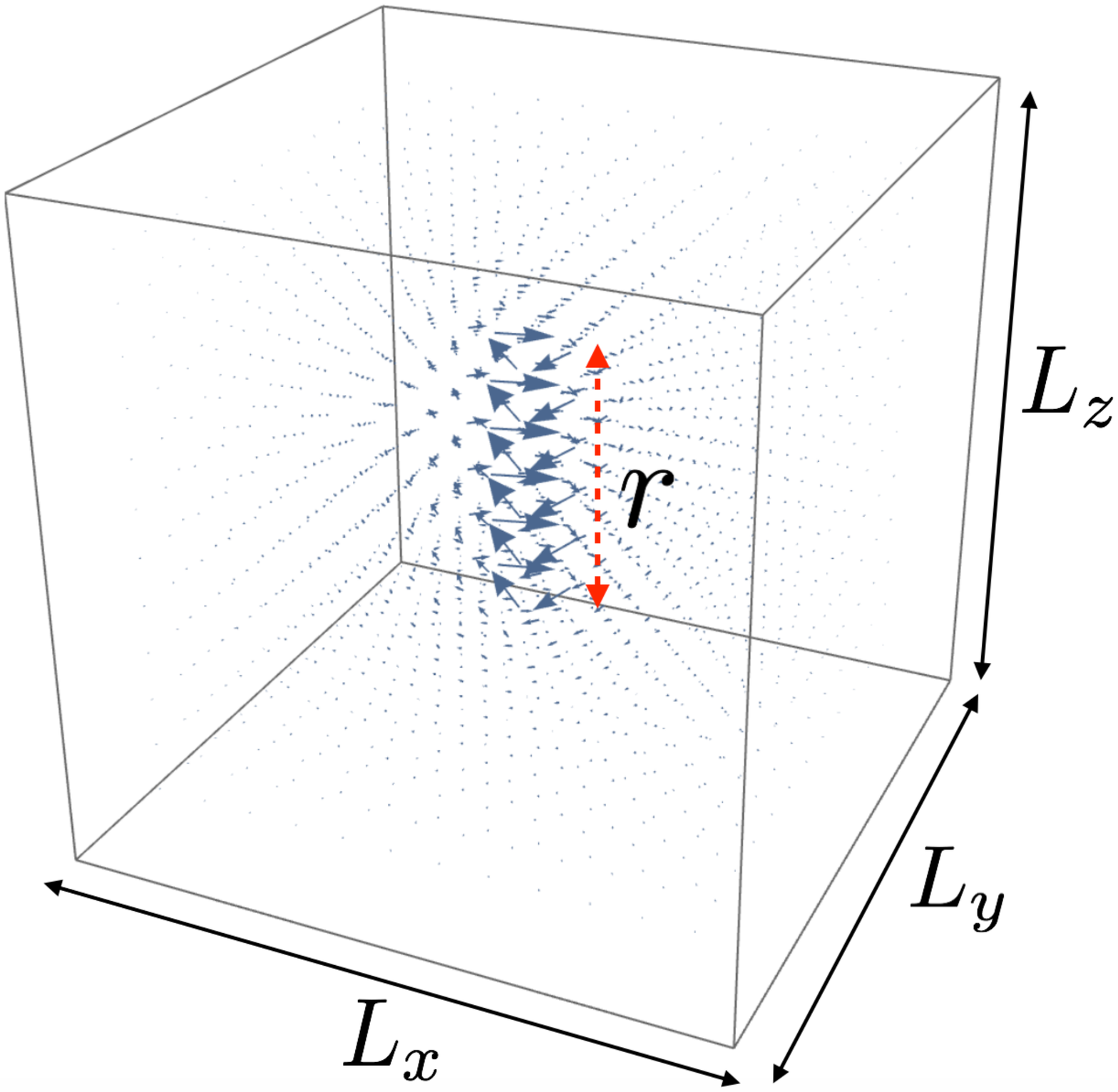}
\caption{The background field ${\cal A}^{Q\overline{Q}}$ from
monopole-antimonopole pair (the blue arrows) in a periodic $L_x\times
L_y\times L_z$ lattice box.  The distance between monopole and
antimonopole is $r$, and they are separated along the $z$-direction.
}
\eef{monopolediag}

\section{Hybrid Monte Carlo for XY model}

We use hybrid Monte Carlo (HMC)~\cite{Duane:1987de} global updates to simulate the XY
model.  Below, we give the HMC force calculation for the general
lattice superconductor model with non-zero $e$, of which the XY
model corresponds to $e=0$.  For HMC, we introduce the auxiliary
momenta $\Pi(x)$ conjugate to $\theta(x)$ and $\pi_\mu(x)$ conjugate
to $a_\mu(x)$. For the fictitious Hamiltonian ${\cal H}$,
\begin{equation}
{\cal H} = \frac{1}{2}\sum_{x}\Pi^2(x) +  \frac{1}{2}\sum_{x,\mu}\pi_\mu^2(x) + S_{XY}(\zeta {\cal A}^{1\overline{1}}).
\end{equation}
where $S_{XY}$ is the action in \eqn{sxy} with the replacement $
{\cal A}^{Q\overline{Q}} \to \zeta {\cal A}^{1\overline{1}}$ in
order to find $W(\zeta)$.  The molecular dynamics evolution through
Monte Carlo time $\tau$ is
\begin{eqnarray}
\frac{d\Pi(x)}{d\tau}&=&-\frac{\partial S}{\partial \theta(x)};\qquad \frac{d\pi_\mu(x)}{d\tau}=-\frac{\partial S}{\partial a_\mu(x)},\cr
\frac{d\theta(x)}{d\tau}&=&\Pi(x);\qquad \frac{d a_\mu(x)}{d\tau}=\pi_\mu(x).
\end{eqnarray}
The explicit expressions  are
\begin{eqnarray}
\frac{d\Pi(x)}{d\tau} &=& \beta\sum_{\mu=1}^3\big{[} \sin\left(\nabla_\mu\theta(x)+e a_\mu(x)+\zeta{\cal A}^{1\overline{1}}_\mu(x)\right) \cr && 
- \sin\left(\nabla_\mu\theta(x-\hat\mu)+e a_\mu(x-\hat\mu)+\zeta {\cal A}^{1\overline{1}}_\mu(x-\hat\mu) \right)\big{]},\cr
\frac{d\pi_\mu(x)}{d\tau} &=& -\frac{1}{2}\sum_{\nu\ne\mu}\big{[} a_\nu(x+\hat\mu)-a_\mu(x+\hat\nu)-a_\nu(x)\cr
&&+a_\nu(x-\nu)-a_\mu(x-\nu)-a_\nu(x-\hat\nu+\hat\mu)\big{]}\cr
 & & - e \beta \sin\left(\nabla_\mu\theta(x)+e a_\mu(x)+\zeta A_\mu(x)\right).
\end{eqnarray}

\bef
\centering
\includegraphics[scale=0.65]{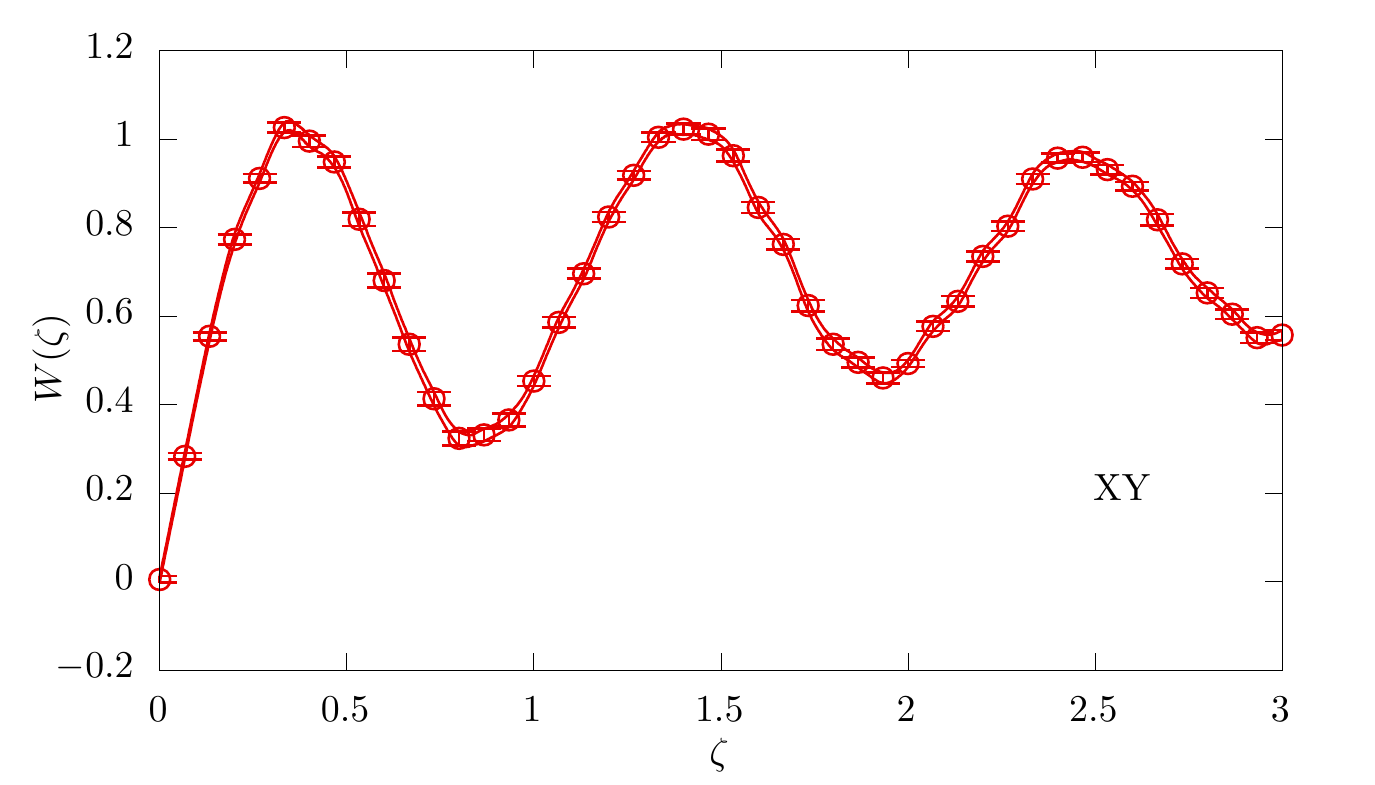}
\includegraphics[scale=0.65]{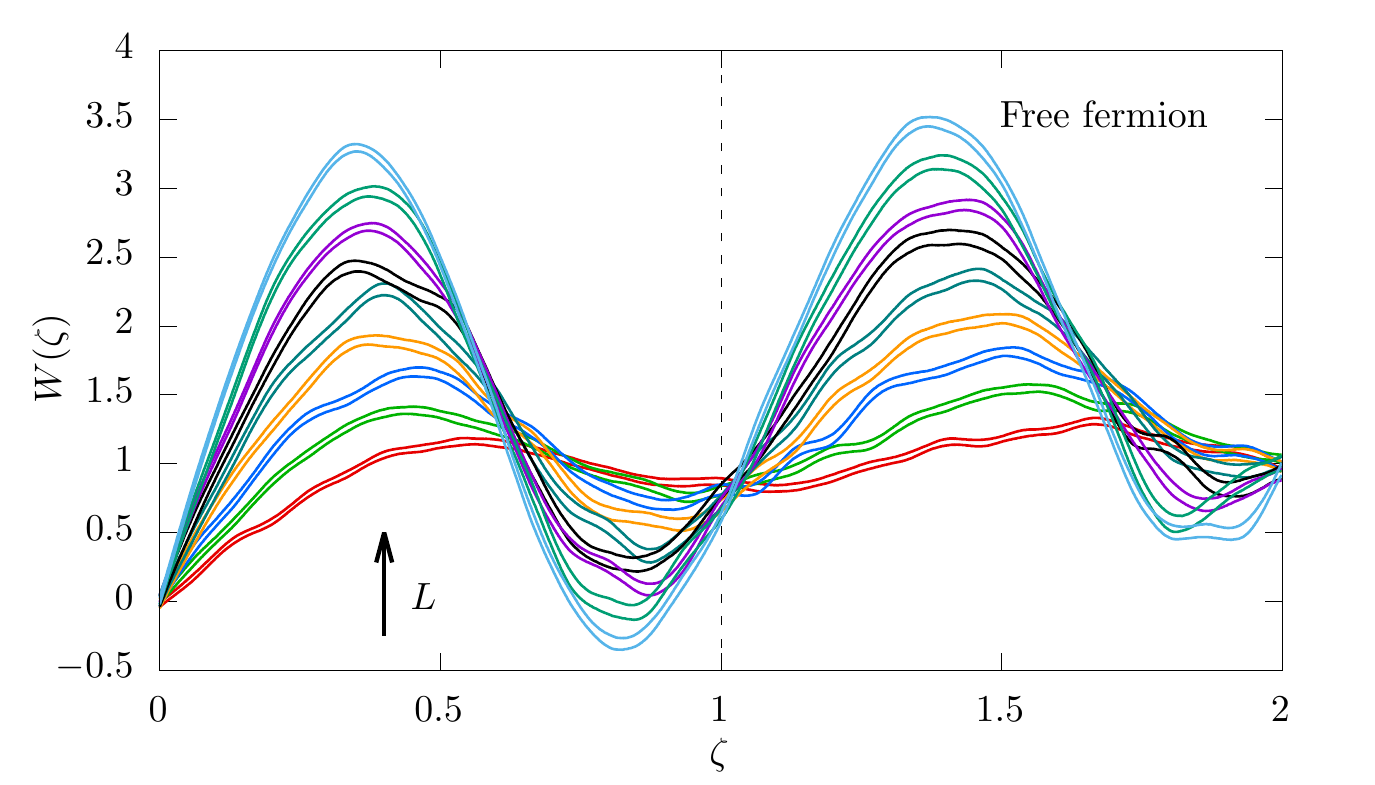}
\includegraphics[scale=0.65]{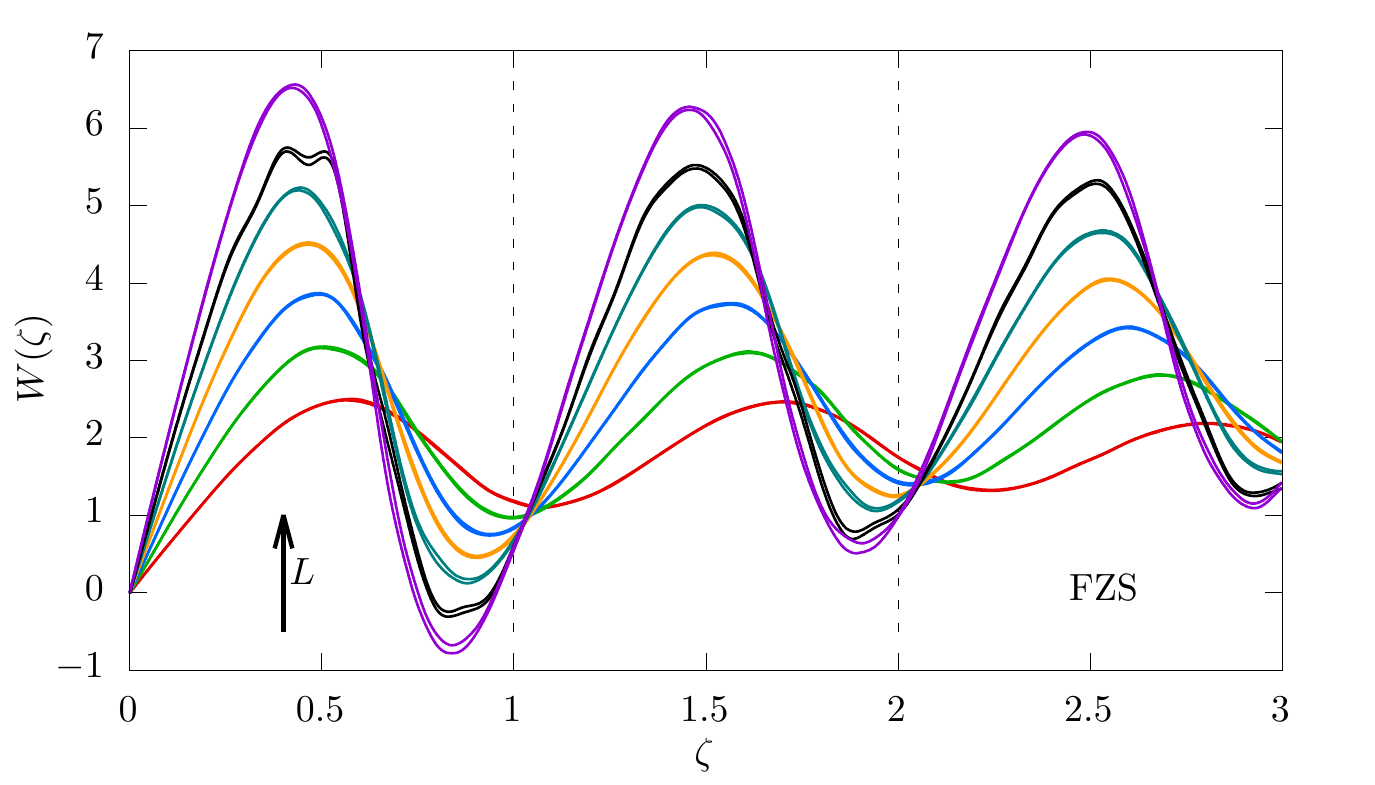}
\caption{Top panel shows $W(\zeta)$ for the critical XY model at
$r/L=1/4$ on $L=20$ lattice.  The measurements of $W(\zeta)$ are the
red circles, while the red band is the 1-$\sigma$ error band from
the cubic spline interpolation. The middle panel shows $W(\zeta)$
for free Wilson-Dirac fermion and the bottom one for the critical
FZS model. The different color bands correspond to different $L$
at fixed $r/L=1/4$.  Along the direction of the arrow, the values
of $L$ for free fermion are $L=12,16,20,24,28,32,36,40,44$ respectively.
For the FZS model, it is $L=12,16,20,24,28,32,36$ respectively. }
\eef{spline}

\section{Determination of $W(\zeta)$}
The definition of $W$ is
\beq
W(\zeta; r) \equiv \frac{-1}{Z(\zeta {\cal A}^{1\overline{1}})}\frac{\partial Z(\zeta {\cal A}^{1\overline{1}})}{\partial \zeta}.
\eeq{wdef}
The right-hand side can be expressed as an ensemble average of
quantities evaluated in the simulation with actions $S(\zeta{\cal
A}^{1\overline{1}})$. Denoting such ensemble averages as
$\langle\ldots\rangle_\zeta$, the expression for $W$ in the XY model
is
\beq
W(\zeta) = \beta \bigg{\langle}\sum_x\sum_{\mu=1}^3 A^{1\overline{1}}_\mu(x)\sin\big{(}\nabla_\mu\theta(x)
+\zeta A^{1\overline{1}}_\mu(x)\big{)}\bigg{\rangle}_\zeta.
\eeq{xyw}
For the FZS model, 
\beqa
W(\zeta) &=& \frac{4\pi^2}{e^2} \sum_x\sum_{\mu>\nu=1}^3\frac{\tilde{\cal{F}}^{1\overline{1}}_{\mu\nu}(x)}{2\pi} \big{(}\nabla_\mu
n_\nu(x)-\nabla_\nu n_\mu(x)\cr
&&-\zeta \frac{\tilde{\cal{F}}^{1\overline{1}}_{\mu\nu}(x)}{2\pi}\big{)}.
\eeqa{fzsw}
Now for the case of free Wilson-Dirac fermion. To avoid the trivial
zero mode in free field theory, we apply anti-periodic boundary
condition in the $z$-direction. The Dirac operator is
\beqa
\slashed{D}_W(x,y) &=&(3-M_W)\delta_{x,y}+\frac{1}{2}\sum_{k=1}^3\big{\{} (\sigma_k+1)U_\mu(x)\delta_{x+\hat k,y}\cr
&& +(1-\sigma_k) U^*_\mu(x-\hat k)\delta_{x-\hat k,y}\big{\}},
\eeqa{wildirac}
where $\sigma_k$ are Pauli matrices, $U_k(x)=e^{i\zeta \tilde{{\cal
A}}^{1\overline{1}}_k(x)}$, and $M_W$ is the Wilson mass which we
tune such that the second smallest eigenvalue of
$\slashed{D}_W^\dagger\slashed{D}_W$ is minimized as a function of
$M_W$ on ${\cal A}^{1\overline{1}}$ background.  Taking the derivative
of $F(\zeta A^{1\overline{1}})=-\log\det\slashed{D}_W$,
\beqa
W(\zeta)&=&-\frac{1}{\det\slashed{D}_W}\frac{\partial}{\partial\zeta}\det\slashed{D}_W,\cr
&=& -{\rm Tr}\left(\slashed{D}_W^{-1}\frac{\partial\slashed{D}_W}{\partial\zeta}\right).
\eeqa{wferm}
The explicit expression for the derivative is
\beqa
&&\cr
&&\frac{\partial}{\partial\zeta}\slashed{D}_W(x,y)=\frac{i}{2}\sum_{k=1}^3\big{\{} (\sigma_k+1)A^{1\overline{1}}_k(x)U_\mu(x)\delta_{x+\hat k,y}\cr
&&-(1-\sigma_k)A^{1\overline{1}}_k(x-\hat k) U^*_\mu(x-\hat k)\delta_{x-\hat k,y}\big{\}}.
\eeqa{wdderiv}
Using these expressions, we determine the trace stochastically using
$N_v\approx 10^4$ Gaussian random vectors $R_i$ satisfying
$\overline{R^{*a}_i R_j^b}=\delta_{i,j}\delta_{a,b}$:
\beq
W(\zeta)=-\frac{1}{N_v}\sum_{i=1}^{N_v}\left\{ R^\dagger_i \slashed{D}_W^{-1}\frac{\partial\slashed{D}_W}{\partial\zeta} R_i\right\}.
\eeq{randvec} 
We used 48 different values of $\zeta$ from 0 to 3 in the case of
XY and FZS models, and up to 2 for free fermion due to the extra
computation with the fermion inversion. We interpolated the actual
Monte Carlo data for $W(\zeta)$ using cubic-spline and integrated
the spline to get the free energy. In the top panel of \fgn{spline},
we show the data as circles and the cubic spline interpolation of
this data as the red, 1-$\sigma$ error band.  The middle and bottom
panels of \fgn{spline} show the behavior of $W(\zeta)$ for free
Wilson-Dirac fermion and critical FZS model respectively.

\end{document}